\documentclass[onecolumn,aps,prd,showpacs,floatfix]{revtex4}

\usepackage{amsmath}
\usepackage{amssymb}
\usepackage{color}
\usepackage{graphicx}
\usepackage{cancel}
\usepackage{bbm}
\usepackage{bm}
\usepackage{maybemath}

\newcommand{\dd}{\mathrm{d}}
\newcommand{\ee}{\mathrm{e}}
\newcommand{\ii}{\mathrm{i}}

\newcommand{\calC}{\mathcal{C}}
\newcommand{\calL}{\mathcal{L}}
\newcommand{\calP}{\mathcal{P}}
\newcommand{\calT}{\mathcal{T}}

\newcommand{\rmT}{\mathrm{T}}

\newcommand{\plus}{{\mbox{{\bf{\tiny +}}}}}

\begin{document}

\title{Pseudo--Hermitian
Quantum Dynamics of Tachyonic Spin--$\maybebm{1/2}$ Particles}

\newcommand{\addrMST}{Department of Physics,
Missouri University of Science and Technology,
Rolla, Missouri 65409-0640, USA}

\author{U. D. Jentschura}
\affiliation{\addrMST}

\author{B. J. Wundt}
\affiliation{\addrMST}

\begin{abstract}
We investigate the spinor solutions, the spectrum and the symmetry properties 
of a matrix-valued wave equation whose plane-wave
solutions satisfy the superluminal (tachyonic)
dispersion relation $E^2 = \vec p^{\,2} -
m^2$, where $E$ is the energy, $\vec p$ is the
spatial momentum, and $m$ is the mass of
the particle. The equation 
reads $(\ii \gamma^\mu \, \partial_\mu - \gamma^5 \, m) \psi = 0$,
where $\gamma^5$ is the fifth current. The tachyonic equation is 
shown to be $\mathcal{CP}$ invariant, and $\mathcal T$ invariant. 
The tachyonic Hamiltonian $H_5 = \vec \alpha \cdot \vec p + \beta \, \gamma^5 \, m$
breaks parity and is non--Hermitian but fulfills the 
pseudo--Hermitian property $H_5(\vec r) = P \, H^\plus_5(-\vec r) \, P^{-1} = 
\calP \, H^\plus_5(\vec r) \, \calP^{-1}$, 
where $P$ is the parity matrix and $\calP$ is the full parity transformation.
The energy eigenvalues and eigenvectors describe a continuous spectrum of 
plane-wave solutions (which correspond to real eigenvalues for $|\vec p| \geq m$) and 
evanescent waves, which constitute resonances and antiresonances 
with complex-conjugate pairs of resonance eigenvalues (for $|\vec p| < m$) .
In view of additional algebraic properties of the Hamiltonian 
which supplement the pseudo-Hermiticity, the existence of a resonance energy eigenvalues $E$
implies that $E^*$, $-E$, and $-E^*$ also constitute resonance energies of $H_5$.\\[2ex]
{\bf Accepted for publication by J.~Phys.~A: Math.~Theor. for the special issue on\\
``Quantum Physics with non--Hermitian Operators''}
\end{abstract}


\maketitle

%
%
\section{Introduction} 

Extensions of the Dirac equation have been discussed in great numbers in the
scientific literature, 
including field-theoretical formulations~\cite{JeWu2012}. 
Notably, there have been efforts to incorporate
space-time curvature into the equation~\cite{GaHuLi2006},
and, to enhance the equation
for spin-$1$ particles such as the photon, and to map the Maxwell equations
onto Dirac-type equations~\cite{Mo2010}. Here, we follow a different route and analyze
whether it is meaningful and consistent to 
modify the mass term in the free Dirac equation so that the
energy-momentum dispersion relation changes from $E^2 = \vec p^{\,2} + m^2$ to $E^2
= \vec p^{\,2} - m^2$.  The latter dispersion relation would be relevant for
superluminal particles, which travel faster than the speed of light. 
The tachyonic spin--$1/2$ equation reads 
\begin{equation}
(\ii \gamma^\mu \, \partial_\mu - \gamma^5 \, m) \; \psi(x) = 0 \,,
\end{equation}
where the $\gamma^\mu$ are the Dirac matrices, 
$\partial_\mu = \partial/\partial x^\mu$ is the derivative
with respect to space-time coordinates, and 
$\gamma^5 = \ii \, \gamma^0 \, \gamma^1 \, \gamma^2 \, \gamma^3$
is the fifth current [here, $x = (t, \vec r)$ denotes a space-time 
four vector].
Because of the occurrence of the Dirac matrices, we propose the 
name `tachyonic Dirac equation'.
The corresponding Lagrangian density reads
\begin{equation}
\calL = \frac{\ii}{2} 
\overline\psi \, \gamma^5 \, \overleftrightarrow{\cancel{\partial}} \psi
- m \, \overline\psi \, \psi 
= \frac{\ii}{2} \left( \overline\psi \, \gamma^5 \, \gamma^\mu (\partial_\mu \psi) -
(\partial_\mu \overline\psi) \, \gamma^5 \, \gamma^\mu \psi   \right) - 
m \, \overline\psi \, \psi \,,
\end{equation}
where $\overline \psi = \psi^\plus \, \gamma^0$ is the Dirac adjoint.
It is perhaps not immediately obvious that
the $\gamma^5$ matrix has to be placed with the current 
matrix $\gamma^\mu$ in the Lagrangian; otherwise an 
inconsistent equation is obtained for $\overline\psi$ 
upon variation with respect to $\psi$.

Indeed, the tachyonic Dirac equation has been indicated by Chodos, Hauser and
Kostelecky in Refs.~\cite{ChHaKo1985} in the equivalent form $(\ii \gamma^5 \,
\gamma^\mu \, \partial_\mu - m) \psi = 0$, where the 
current is modified instead of the mass term (see also
Refs.~\cite{ChKoPoGa1992,ChKo1994}). More recently, the noncovariant
(Hamiltonian) form of the equation has been indicated in Refs.~\cite{Ch2000,Ch2002},
with a preliminary ansatz for its solution.  Very recently, the equation
has been used in the context of the Gross-Neveu model in Ref.~\cite{Oi2011}.
However, neither the complete set of solutions, nor the spectrum, nor the 
symmetries of the equation have been 
fully investigated in the literature to the best 
of our knowledge. 

An important observation made in the current work concerns the 
Hamiltonian 
\begin{equation}
H_5 = \vec \alpha \cdot \vec p + \beta \, \gamma^5 \, m
\end{equation}
which is obtained from the tachyonic Dirac equation. Here,
$\vec \alpha = \gamma^0 \, \vec \gamma$ and $\beta = \gamma^0$,
with the identification of the Dirac matrices discussed below
in Eq.~\eqref{diracrep}.
The Hamiltonian $H_5$ breaks parity and is non--Hermitian but fulfills the
pseudo--Hermitian equation 
\begin{equation}
H_5(\vec r) = P \, H^\plus_5(-\vec r) \, P^{-1} 
            = \calP \, H^\plus_5(\vec r) \, \calP^{-1},
\end{equation}
where $P$ is the parity matrix and $\calP$ is the full parity transformation.
While the scalar product induced by $\calP$ is not positive semi-definite,
a few essential properties of the Hamiltonian persist 
under the pseudo--Hermiticity and help 
to characterize the spectrum.
We thus extend recent work on non-Hermitian, 
but pseudo-Hermitian or $\calP \calT$-symmetric 
Hamiltonians~\cite{BeBo1998,BeDu1999,BeBoMe1999,BeWe2001,BeBrJo2002,
Mo2002i,Mo2002ii,Mo2002iii,Mo2003npb,%
JeSuZJ2009prl,JeSuZJ2010}
which have also been applied in the field-theoretical
context~\cite{BeJoRi2005,JSMa2009prep1,JSMa2009prep2}.

We proceed as follows:
In Sec.~\ref{CPT}, we analyze the transformation properties
of the covariant formulation of the tachyonic
Dirac equation under charge conjugation,
parity transformation and time reversal. 
The propagator and the explicit solutions of
the tachyonic Dirac equation are analyzed in 
Sec.~\ref{sol}. 
Finally, conclusions are drawn in Sec.~\ref{conclu}.
Units with $\hbar = c = \epsilon_0 = 1$ are used throughout 
the paper.

%
%
\section{Discrete Symmetries: $\bm{\calC}$, $\bm{\calP}$, and $\bm{\calT}$}
\label{CPT}

%
%
\subsection{Charge Conjugation}
\label{SecFifthC}

We start from
\begin{equation}
\left[ \gamma^\mu \left( \ii \partial_\mu - e \, A_\mu \right) - 
\gamma^5 m \right] \, \psi = 0 \,,
\end{equation}
where the $\gamma^\mu$ are the Dirac $\gamma$ matrices, which
we use in the Dirac representation,
\begin{align}
\label{diracrep}
\gamma^0 =& \; \left( \begin{array}{cc} \mathbbm{1}_{2\times 2} & 0 \\
0 & -\mathbbm{1}_{2\times 2} \\
\end{array} \right) \,,
\quad
\vec\gamma = \left( \begin{array}{cc} 0 & \vec\sigma \\ -\vec\sigma & 0  \\
\end{array} \right) \,,
\quad
\gamma^5 = \left( \begin{array}{cc} 0 & \mathbbm{1}_{2\times 2} \\
\mathbbm{1}_{2\times 2} & 0  \\
\end{array} \right) \,.
\end{align}
The Dirac $\gamma^\mu$ matrices fulfill the following
anticommutator relations,
\begin{equation}
\left\{ \gamma^\mu, \gamma^\nu \right\} = 2 \, g^{\mu\nu} \,,
\end{equation}
where $g^{\mu\nu} = {\rm diag}(1,-1,-1,-1)$
is the space-time metric.
Transposition and complex conjugation leaves $\gamma^5$ invariant,
\begin{equation}
\psi^\plus \left( \left( \gamma^\mu \right)^\plus
 \left( -\ii \overleftarrow{\partial}_\mu - e \, A_\mu \right) 
- \gamma^5 \, m \right) = 0 \,,
\end{equation}
where the partial derivative acts on the left.
The Dirac adjoint is introduced as
\begin{equation}
\left( \psi^\plus \gamma^0 \right)
\gamma^0 \left( \left( \gamma^\mu \right)^\plus
 \left( -\ii \overleftarrow{\partial}_\mu - e \, A_\mu \right)
 - \gamma^5 m \right) \gamma^0 = 0 \,.
\end{equation}
Using the  identities
$\gamma^0 \left( \gamma^\mu \right)^\plus \gamma^0 =  \gamma^\mu$
and $\gamma^0 \gamma^5 \gamma^0 = -\gamma^5$,
one writes
\begin{equation}
\bar\psi 
\left( \gamma^\mu \left( -\ii \overleftarrow{\partial}_\mu - e \, A_\mu \right) + 
\gamma^5 \,  m \right) = 0 \,.
\end{equation}
A further transposition leads to 
\begin{equation}
\left( \left( \gamma^\mu \right)^{\rm T} 
\left( -\ii \partial_\mu - e \, A_\mu \right) + \gamma^5 m \right) 
\bar\psi^{\rm T} = 0 \,.
\end{equation}
Introducing the charge conjugation matrix $C$
with the property
\begin{equation}
C  \left( \gamma^\mu \right)^{\rm T} C^{-1} = -\gamma^\mu \,,
\end{equation}
the charge conjugation of the fifth current
is calculated symbolically as
\begin{align}
& C \, \left( \gamma^5 \right)^{\rm T} \, C^{-1} =
C \, \ii 
\left( \gamma^3 \right)^{\rm T} \, 
\left( \gamma^2 \right)^{\rm T} \, 
\left( \gamma^1 \right)^{\rm T} \, 
\left( \gamma^0 \right)^{\rm T} \, 
C^{-1} 
\nonumber\\[0.77ex]
& = \ii
C \left( \gamma^3 \right)^{\rm T}  C^{-1} \,
C \left( \gamma^2 \right)^{\rm T}  C^{-1} \,
C \left( \gamma^1 \right)^{\rm T} C^{-1} \,
C \left( \gamma^0 \right)^{\rm T}  C^{-1}
\nonumber\\[0.77ex]
& = \ii \left( -\gamma^3 \right) \;
\left( -\gamma^2 \right) \;
\left( -\gamma^1 \right) \;
\left( -\gamma^0 \right) 
\nonumber\\[0.77ex]
& = \ii \, \gamma^3 \, \gamma^2 \, \gamma^1 \, \gamma^0 
= \ii \, \gamma^0 \, \gamma^1 \, \gamma^2 \, \gamma^3 = \gamma^5  \,.
\end{align}
Therefore, $\psi^\calC = C \, \bar\psi^{\rm T}$ fulfills
the charge conjugate tachyonic Dirac equation,
\begin{equation}
\label{FifthC}
\left[ \gamma^\mu 
\left( \ii \, \partial_\mu + e \, A_\mu \right) + \gamma^5 m \right] \;
\psi^\calC = 0 \,.
\end{equation}
As a result of the calculation of the Dirac adjoint
($\gamma^5 \to \gamma^0 \, \gamma^5 \, \gamma^0 = - \gamma^5$)
and the charge conjugation using the $C$ matrix, the term with the $\gamma^5$ matrix 
has changed sign. In contrast to the ordinary Dirac 
equation~\cite{ItZu1980}, the tachyonic equation is not $\calC$ invariant.

%
%
\subsection{Parity}
\label{SecFifthP}

Again, we start from
\begin{equation}
\left( \ii \gamma^\mu \partial_\mu - \gamma^5 m \right) \, \psi(x) =
\left( 
\ii \gamma^0 \partial_0 +
\ii \gamma^i \partial_i - \gamma^5 m \right) \, \psi(x) = 0 \,.
\end{equation}
Under parity, $x \to x_P = (t, - \vec r)$ and so 
\begin{equation}
\left(
\ii \gamma^0 \, \partial_0 -
\ii \gamma^i \, \partial_i - \gamma^5 \, m \right) \, \psi(x_P) = 0 \,.
\end{equation}
One introduces a parity transformation matrix $P$,
where $P = \gamma^0$ in the Dirac representation, with the property
\begin{equation}
P \gamma^0 P^{-1} = \gamma^0 \,,
\quad
P \gamma^i P^{-1} = -\gamma^i \,,
\quad
P \gamma^5 P^{-1} = -\gamma^5 \,,
\end{equation}
where again $i=1,2,3$ is spatial. So, $\psi^\calP(x) = P \, \psi(x_P)$
fulfills the parity transformed tachyonic Dirac equation,
\begin{equation}
P \, \left( \ii \gamma^0 \partial_0 -
\ii \gamma^i \partial_i - \gamma^5 \, m \right) \, P^{-1}\,\psi^\calP(x) = 0 
\end{equation}
which can be rewritten as
\begin{equation}
\left( \ii \gamma^0 \partial_0 +
\ii \gamma^i \partial_i + \gamma^5 \, m \right) \, \psi^\calP(x) = 0 \,.
\end{equation}
In covariant notation, we have 
\begin{equation}
\label{FifthP}
\left( \ii \gamma^\mu \, \partial_\mu
+ \gamma^5 m \right) \;
\psi^\calP(x) = 0 \,.
\end{equation}
The tachyonic Dirac equation is not parity invariant.

%
%
\subsection{Time Reversal}
\label{SecFifthT}

Starting from the tachyonic Dirac equation,
\begin{equation}
\left( \ii \gamma^\mu \partial_\mu - \gamma^5 m \right) \, \psi = 0 \,,
\end{equation}
transposition and complex conjugation leads to
\begin{equation}
\psi^\plus \left( -\ii \left( \gamma^\mu \right)^\plus 
\overleftarrow{\partial}_\mu  
- \gamma^5 m \right) = 0 \,,
\end{equation}
where again the differential operator 
$\overleftarrow{\partial}_\mu$ acts to the left.
The insertion of the $\gamma^0$ matrix
ensures Lorentz covariance and leads to
\begin{equation}
\left( \psi^\plus \gamma^0 \right)
\gamma^0 \left( -\ii \left( \gamma^\mu \right)^\plus 
\overleftarrow{\partial}_\mu - \gamma^5 \, m \right) \gamma^0 = 0 \,,
\end{equation}
which is equivalent to
\begin{equation}
\bar\psi 
\left( -\ii \gamma^\mu \overleftarrow{\partial}_\mu
+ \gamma^5 \, m \right) = 0 \,.
\end{equation}
We here use the fact that a proper formulation of 
time reversal must respect the fact that the 
in and out states interchange under $\calT$.
We thus have to formulate the time reversal in terms of the 
adjoint of the Dirac equation 
(see Sec.~4.4.2 of Ref.~\cite{HeLect}).
Again, we transpose and write
\begin{equation}
\left( -\ii \, \left( \gamma^\mu \right)^{\rm T} 
\partial_\mu + \gamma^5 \, m \right) \bar\psi^{\rm T} = 0 \,.
\end{equation}
We introduce the time reversal matrix $T$, with the properties
($\calT$ means time reversal, $T$ is the time reversal matrix,
and $\rm T$ denotes the transpose),
\begin{equation}
T \left( \gamma^0 \right)^{\rm T} T^{-1} = \gamma^0 \,,
\qquad
T \left( \gamma^i \right)^{\rm T} T^{-1} = -\gamma^i \,.
\end{equation}
The time reversal of the fifth current is calculated as
\begin{align}
& T \, \left( \gamma^5 \right)^{\rm T} \, T^{-1} =
T \, \ii 
\left( \gamma^3 \right)^{\rm T} \, 
\left( \gamma^2 \right)^{\rm T} \, 
\left( \gamma^1 \right)^{\rm T} \, 
\left( \gamma^0 \right)^{\rm T} \, 
T^{-1} 
\nonumber\\[0.77ex]
& = \ii
T \left( \gamma^3 \right)^{\rm T} T^{-1} \,
T \left( \gamma^2 \right)^{\rm T} T^{-1} \,
T \left( \gamma^1 \right)^{\rm T} T^{-1} \,
T \left( \gamma^0 \right)^{\rm T} T^{-1} 
\nonumber\\[0.77ex]
& = \ii \left( -\gamma^3 \right) \;
\left( -\gamma^2 \right) \;
\left( -\gamma^1 \right) \;
\left( +\gamma^0 \right) 
\nonumber\\[0.77ex]
& = -\ii \gamma^3 \gamma^2 \gamma^1 \gamma^0 
= -\ii \gamma^0 \gamma^1 \gamma^2 \gamma^3 = -\gamma^5  \,.
\end{align}
The time-reversed tachyonic Dirac equation thus reads as
\begin{equation}
\label{FifthT}
\left( \ii \gamma^\mu \, \partial_\mu - \gamma^5 \, m \right) 
\psi^\calT(x) = 0 \,,
\end{equation}
where $\psi^\calT(x) = T \, \bar\psi^{\rm T}(x_T)$
with $x = (-t, \vec r)$.
As a result of the replacements $\gamma^5 \to \gamma^0 \gamma^5 \gamma^0$ and 
$\gamma^5 \to T \gamma^5 T^{-1}$,
the tachyonic Dirac equation is seen to be ${\mathcal T}$ invariant.

%
%
\subsection{Overview of the Symmetry Properties}

A brief summary is in order.
The charge conjugation matrix fulfills~\cite{ItZu1980,PeSc1995}
\begin{equation}
C \left( \gamma^\mu \right)^{\rm T} C^{-1} = -\gamma^\mu \,.
\end{equation}
In the Dirac representation, 
a possible choice is $C = \ii \, \gamma^2 \, \gamma^0$.
The parity transformation fulfills
\begin{equation}
P \gamma^0 P^{-1} = \gamma^0 \,,
\qquad
P \gamma^i P^{-1} = -\gamma^i \,.
\end{equation}
In the Dirac representation, we may choose $P = \gamma^0$.
The time reversal operation fulfills
\begin{equation}
T \left( \gamma^0 \right)^{\rm T} T^{-1} = \gamma^0 \,,
\qquad
T \left( \gamma^i \right)^{\rm T} T^{-1} = -\gamma^i \,.
\end{equation}
In the Dirac representation of the 
$\gamma$ matrices, one can choose $T = \ii \, \gamma^2 \, \gamma^5$.
The matrices $C$, $P$ and $T$ fulfill the relations 
$C^{\rm T} = C^{-1}$,
$P^{\rm T} = P^{-1}$,
$T^{\rm T} = T^{-1}$,
where by $\rmT$ we denote the transpose.
A representation of $\calC\calP\calT$ is given by
\begin{equation}
C P T = \gamma^5 \,.
\end{equation}
This implies that 
\begin{equation}
(CPT) \gamma^\mu (CPT)^{-1} = -\gamma^\mu \,,
\end{equation}
a relation which has been given on page~239 of the 
textbook~\cite{BBBB1975} and in Eq.~(40.48) of Ref.~\cite{Sr2007}.
We also recall the
transformation properties of $\gamma^5$ just derived,
\begin{equation}
\begin{split}
C \left( \gamma^5 \right)^{\rm T} C^{-1} = \gamma^5 \,,
\\
P \gamma^5 P^{-1} = -\gamma^5 \,,
\\
T \left( \gamma^5 \right)^{\rm T} T^{-1} = -\gamma^5 \,.
\end{split}
\end{equation}
We note that the transformation property for $T$ depends on our 
precise formulation of the time reversal operation 
as detailed in Sec~\ref{SecFifthT}.
According to Eqs.~\eqref{FifthC},~\eqref{FifthP}, and~\eqref{FifthT},
the tachyonic 
Dirac equation is separately $\calC \calP$ invariant, and $\calT$ invariant.

%
%
\section{Solutions of the Tachyonic Equation}
\label{sol}

%
%
\subsection{Plane--Wave Solutions and Chirality}
\label{solutions}

We start from the eigenfunctions of the 
operator $\vec\sigma \cdot \vec k$, which are given by
\begin{subequations}
\begin{align}
a_+(\vec k) =& \; \left( \begin{array}{c} 
\cos\left(\frac{\theta}{2}\right) \\[0.77ex]
\sin\left(\frac{\theta}{2}\right) \, \ee^{\ii \, \varphi} \\
\end{array} \right) \,,
\qquad
a_-(\vec k) = \left( \begin{array}{c} 
-\sin\left(\frac{\theta}{2}\right) \, \ee^{-\ii \, \varphi} \\[0.77ex]
\cos\left(\frac{\theta}{2}\right) \\
\end{array} \right) \,,
\end{align}
and fulfill 
\begin{align}
\frac{\vec \sigma \cdot \vec k}{|\vec k|} \, a_\pm(\vec k) =& \; \pm a_\pm(\vec k) \,,
\\[2ex]
\sum_\pm a_\pm(\vec k) \otimes a^\plus_\pm(\vec k) =& \; \mathbbm{1}_2\,,
\qquad
\sum_\pm (-1)^\pm \; a_\pm(\vec k) \otimes a^\plus_\pm(\vec k) = 
\frac{\vec \sigma \cdot \vec k}{|\vec k|} \,,
\end{align}
\end{subequations}
where $\theta$ and $\varphi$ are the polar and azimuthal angles
of the wave vector $\vec k$.
The normalized positive-energy chirality and helicity eigenstates of the 
massless Dirac equation are
\begin{equation}
u_+(\vec k) = 
\frac{1}{\sqrt{2}} 
\left( \begin{array}{c}
a_+(\vec k) \\[0.77ex]
a_+(\vec k) \\
\end{array} \right) \,,
\quad
u_-(\vec k) = 
\frac{1}{\sqrt{2}} 
\left( \begin{array}{c}
a_-(\vec k) \\[0.77ex]
-a_-(\vec k) \\
\end{array} \right) \,.
\end{equation}
These eigenstates immediately lead to 
plane-wave solutions of the massless Dirac equation,
which are also eigenstates of the chirality $\gamma^5$ 
and of the helicity $\vec\Sigma \cdot \vec p/|\vec p|$, where
$\vec \Sigma$ is the vector of ($4\times 4$)-spin matrices,
\begin{subequations}
\begin{align}
\psi(x) =& \; u_\pm(\vec k) \, \ee^{-\ii k \cdot x} \,,
\quad
k = (E, \vec k) \,,
\quad 
E = |\vec k| \,,
\\[0.77ex]
\ii \, \gamma^\mu \, \partial_\mu \psi(x) = & \;
\gamma^\mu \, k_\mu \psi(x) = 0 \,,
\qquad
\ii \, \partial_0 \psi(x) = E \, \psi(x) \,,
\\[0.77ex]
\frac{\vec\Sigma \cdot \vec k}{|\vec k|} \, u_\pm(\vec k) = & \;
\gamma^5 \,  u_\pm(\vec k) = \pm u_\pm(\vec k)\,,
\qquad
\frac{\vec\Sigma \cdot \vec p}{|\vec p|} \, u_\pm(\vec k) \,
\ee^{\ii \, \vec k \cdot \vec r} = \pm u_\pm(\vec k) \, 
\ee^{\ii \, \vec k \cdot \vec r}\,.
\end{align}
\end{subequations}
For the negative-energy solutions of the massless Dirac
equation, we need
\begin{align}
v_+(\vec k) = & \;
\frac{1}{\sqrt{2}} 
\left( \begin{array}{c}
-a_+(\vec k) \\[0.77ex]
-a_+(\vec k) \\
\end{array} \right) \,,
\qquad
v_-(\vec k) =
\frac{1}{\sqrt{2}} 
\left( \begin{array}{c}
-a_-(\vec k) \\[0.77ex]
a_-(\vec k) \\
\end{array} \right) \,.
\end{align}
The negative-energy states fulfill the relations
\begin{subequations}
\begin{align}
\phi(x) =& \; v_\pm(\vec k) \, \ee^{\ii k \cdot x} \,,
\quad
k = (E, \vec k) \,,
\quad
E = |\vec k| \,,
\\[0.77ex]
\ii \, \gamma^\mu \, \partial_\mu \phi(x) = & \;
-\gamma^\mu \, k_\mu \phi(x) = 0 \,,
\qquad
\ii \, \partial_0 \phi(x) = -E \, \phi(x) \,,
\\[0.77ex]
\frac{\vec\Sigma \cdot \vec k}{|\vec k|} \, v_\pm(\vec k) = & \;
\gamma^5 \,  v_\pm(\vec k) = \pm v_\pm(\vec k)\,,
\qquad
\frac{\vec\Sigma \cdot \vec p}{|\vec p|} \, v_\pm(\vec k) \,
\ee^{-\ii \, \vec k \cdot \vec r} = \mp v_\pm(\vec k) \, 
\ee^{-\ii \, \vec k \cdot \vec r}\,.
\end{align}
\end{subequations}
The subscript of the $v$ vectors is chosen according to the 
corresponding eigenvalue of the chirality $\gamma^5$,
i.e., $\gamma^5 \, u_\pm = \pm u_\pm$, and  
$\gamma^5 \, v_\pm = \pm v_\pm$.
For negative-energy solutions, chirality and 
helicity are opposite~\cite{ItZu1980}.
It is worth noting that the set of four states,
\begin{equation}
u_\pm(\vec k) \, \ee^{\ii \vec k \cdot \vec r} \,,
\qquad
v_\pm(-\vec k) \, \ee^{\ii \vec k \cdot \vec r} \,,
\end{equation}
form a complete and orthonormal set of eigenstates of the 
massless Hamiltonian $H_0$,
which is a Hermitian operator,
\begin{subequations}
\begin{align}
H_0 = & \; \vec\alpha \cdot \vec p\,,\qquad
\vec\alpha = \gamma^0 \, \vec \gamma \,, \\[2ex]
H_0 \, u_\pm(\vec k) \, \ee^{\ii \vec k \cdot \vec r} = & \;
|\vec k| \, u_\pm(\vec k) \, \ee^{\ii \vec k \cdot \vec r}\,, 
\qquad
H_0 \, v_\pm(-\vec k) \, \ee^{\ii \vec k \cdot \vec r} = 
-|\vec k| \, v_\pm(-\vec k) \, \ee^{\ii \vec k \cdot \vec r} \,, \\[2ex]
u^\plus_\alpha(\vec k) \, u_\beta(\vec k) =& \; \delta_{\alpha\beta} \,,
\qquad
v^\plus_\alpha(-\vec k) \, v_\beta(-\vec k) =
\delta_{\alpha\beta} \,,
\qquad
u^\plus_\alpha(\vec k) \, v_\beta(-\vec k) = 0 \,, 
\end{align}
\end{subequations}
where $\alpha,\beta = \pm$.
These relations hold in view of 
$a^\plus_+(\vec k) \, a_+(-\vec k) 
= a^\plus_-(\vec k) \, a_-(-\vec k) = 0$.
In view of the relation 
$(\gamma^\mu k_\mu - \gamma^5 \, m)^2 = k^2 + m^2 = E^2 - \vec k^{\,2} + m^2$, 
the generalization of these solutions to the massive
tachyonic Dirac equation is rather straightforward.
We find
\begin{subequations}
\label{UU}
\begin{align}
U_+(\vec k) = & \;
\frac{\gamma^5\,m-\gamma^\mu \, k_\mu}%
{\sqrt{2} \, \sqrt{(E - |\vec k|)^2 + m^2}} \, u_+(\vec k) 
\nonumber\\
=& \;
\left( \begin{array}{c}
\dfrac{m-E+|\vec k|}{\sqrt{2} \, \sqrt{(E - |\vec k|)^2 + m^2}} \; a_+(\vec k) \\[2ex]
\dfrac{m+E-|\vec k|}{\sqrt{2} \, \sqrt{(E - |\vec k|)^2 + m^2}} \; a_+(\vec k) \\
\end{array} \right) 
\end{align}
for positive chirality and
\begin{align}
U_-(\vec k) = & \;
\frac{\gamma^\mu \, k_\mu - \gamma^5\,m}%
{\sqrt{2} \, \sqrt{(E - |\vec k|)^2 + m^2}} \, u_-(\vec k)
\nonumber\\
=& \;
\left( \begin{array}{c}
\dfrac{m+E-|\vec k|}{\sqrt{2} \, \sqrt{(E - |\vec k|)^2 + m^2}} \; a_-(\vec k) \\[2ex]
\dfrac{-m+E-|\vec k|}{\sqrt{2} \, \sqrt{(E - |\vec k|)^2 + m^2}} \; a_-(\vec k) \\
\end{array} \right) 
\end{align}
\end{subequations}
for negative chirality.
Here, $E = \sqrt{\vec k^{2} - m^2}$ with $\vec k^{\,2} > m^2$.
For $m \to 0$, the particle has to be on the mass shell,
therefore $E \to |\vec k|$, and we have the tachyon spinors
approaching their massless limit, i.e.,
$U_+(\vec k) \to u_+(\vec k)$ and
$U_-(\vec k) \to u_-(\vec k)$. The negative-energy 
eigenstates are
\begin{subequations}
\label{VV}
\begin{align}
V_+(\vec k) = & \;
\frac{\gamma^5\,m+\gamma^\mu \, k_\mu}%
{\sqrt{2} \, \sqrt{(E - |\vec k|)^2 + m^2}} \, u_+(\vec k)
\nonumber\\
=& \;
\left( \begin{array}{c}
\dfrac{-m-E+|\vec k|}{\sqrt{2} \, \sqrt{(E - |\vec k|)^2 + m^2}} \; a_+(\vec k) \\[2ex]
\dfrac{-m+E-|\vec k|}{\sqrt{2} \, \sqrt{(E - |\vec k|)^2 + m^2}} \; a_+(\vec k) \\
\end{array} \right) 
\end{align}
for positive chirality (negative helicity) and
\begin{align}
V_-(\vec k) = & \;
\frac{-\gamma^\mu \, k_\mu - \gamma^5\,m}%
{\sqrt{2} \, \sqrt{(E - |\vec k|)^2 + m^2}} \, u_-(\vec k)
\nonumber\\
=& \;
\left( \begin{array}{c}
\dfrac{-m+E-|\vec k|}{\sqrt{2} \, \sqrt{(E - |\vec k|)^2 + m^2}} \; a_-(\vec k) \\[2ex]
\dfrac{m+E-|\vec k|}{\sqrt{2} \, \sqrt{(E - |\vec k|)^2 + m^2}} \; a_-(\vec k) \\
\end{array} \right) 
\end{align}
\end{subequations}
for negative chirality (positive helicity).
For $m \to 0$ and $E \to |\vec k|$, we have
$V_+(\vec k) \to v_+(\vec k)$ and
$V_-(\vec k) \to v_-(\vec k)$. 
These states are normalized, i.e., 
$U^\plus_+(\vec k) \, U_+(\vec k) =
U^\plus_-(\vec k) \, U_-(\vec k) =
V^\plus_+(\vec k) \, V_+(\vec k) =
V^\plus_-(\vec k) \, V_-(\vec k) = 1$.
The corresponding positive-energy solutions
of the massive tachyonic Dirac equation are given as
\begin{subequations}
\begin{align}
& \Psi(x) = U_\pm(\vec k) \, \ee^{-\ii k \cdot x} \,,
\quad
k = (E, \vec k) \,,
\quad
E = \sqrt{\vec k^{2} - m^2} \,,
\\[0.77ex]
& \left( \ii \, \gamma^\mu \, \partial_\mu - \gamma^5 m\right) \Psi(x) = 
(\gamma^\mu \, k_\mu - \gamma^5 m) \Psi(x) = 0 \,,
\\[0.77ex]
& \ii \, \partial_0 \Psi(x) = E \, \Psi(x) \,.
\end{align}
\end{subequations}
The negative-energy solutions are given by
\begin{subequations}
\begin{align}
& \Phi(x) = V_\pm(\vec k) \, \ee^{\ii k \cdot x} \,,
\quad
k = (E, \vec k) \,,
\quad
E = \sqrt{\vec k^2 - m^2} \,,
\\[0.77ex]
& \left( \ii \, \gamma^\mu \, \partial_\mu - \gamma^5 m\right) \Phi(x) =
(-\gamma^\mu \, k_\mu - \gamma^5 m) \Phi(x) = 0 \,,
\\[0.77ex]
& \ii \, \partial_0 \Phi(x) = -E \, \Phi(x) \,,
\end{align}
\end{subequations}
All of the discussed solutions fulfill the 
superluminal dispersion relation
\begin{equation}
\label{dispersion}
E^2 = \vec k^{\,2} - m^2
\end{equation}
and are eigenstates of the tachyonic Dirac
Hamiltonian~\eqref{fifthH}, as given below.

\begin{figure}[t!]
\begin{center}
\begin{minipage}{0.8\linewidth}
\includegraphics[width=0.7\linewidth]{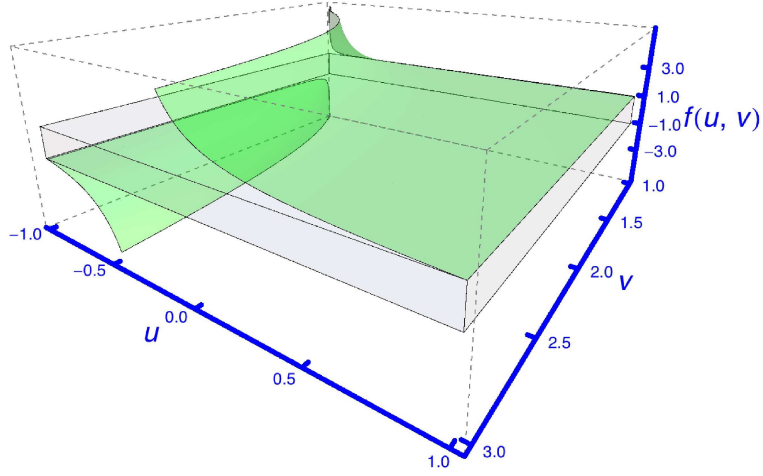}
\caption{\label{fig1} (Color.) Illustration of the 
velocity addition theorem $f(u,v) = (u+v)/(1 + u \, v)$, in 
the superluminal domain with $u \in (-1,1)$ and 
$v \in (1,3)$. The superluminal velocity $v$ remains
superluminal upon transformation from a frame 
which moves at subluminal velocity $u$ with 
respect to the observer; the domain $-1 < f(u,v) < 1$ is excluded,
consistent with Eq.~\eqref{tachdisp}.}
\end{minipage}
\end{center}
\end{figure}

%
%
\subsection{Plane--Wave Solutions and Hamiltonian} 
\label{PTnoncov}

It is instructive to investigate the Hamiltonian form of 
the tachyonic Dirac equation, which reads
\begin{equation}
\left( \ii \, \gamma^\mu \, \frac{\partial}{\partial x^\mu} 
- \gamma^5 \, m \right) \, \psi(x) = 0 \,.
\end{equation}
Multiplication by $\gamma^0$ leads to 
the non-covariant Hamiltonian formulation,
\begin{equation}
H_5 \, \psi(\vec r) = 
\left( \vec \alpha \cdot \vec p + \beta \, \gamma^5 \, m \right) \, \psi(\vec r) = 
E \, \psi(\vec r) \,,
\end{equation}
where $\vec \alpha = \gamma^0 \, \vec \gamma$ and $\beta = \gamma^0$.  
The tachyonic Dirac Hamiltonian obviously reads as
\begin{equation}
\label{fifthH}
H_5 = H_5(\vec r) = 
\vec \alpha \cdot \vec p + \beta \, \gamma^5 \, m  \,,
\end{equation}
where $\vec p = - \ii\vec \nabla$ is the momentum operator.
The Hermitian adjoint is 
\begin{equation}
H_5^\plus(\vec r) =
\vec \alpha \cdot \vec p - \beta \, \gamma^5 \, m  \,,
\end{equation}
where by the superscript $+$ we denote the Hermitian adjungation.
Therefore,
\begin{align}
P \, H_5^\plus(\vec r) \, P^{-1} =& \;
P \, 
\left( \vec \alpha \cdot \vec p - \beta \, \gamma^5 \, m \right) \,
P^{-1} 
\nonumber\\[0.77ex]
=& \; -\vec \alpha \cdot \vec p + \beta \, \gamma^5 \, m \,,
\end{align}
because $\left( \beta \, \gamma^5 \right)^\plus 
= \left(\gamma^5 \right)^\plus \, \left( \gamma^0 \right)^\plus
= \gamma^5 \, \gamma^0 = -\gamma^0 \, \gamma^5$.
Here, $\vec p \to \vec p^{\plus} = \vec p$ because the 
Hermitian adjoint of the momentum operator is 
the momentum operator itself.
It is thus easy to check that the tachyonic Hamiltonian $H_5$ 
is a pseudo-Hermitian 
operator~\cite{Pa1943,Mo2002i,Mo2002ii,Mo2002iii,Mo2003npb},
with the property 
\begin{equation}
\label{pseudoH}
H_5(\vec r) = P \, H_5^+(-\vec r) P^{-1} 
= \calP \, H_5^+(\vec r) \calP^{-1} \,,
\end{equation}
where again $P = \gamma^0$ is the matrix representation of 
parity and $\calP$ is the full parity transformation.
Hamiltonians with this property have been analyzed about seventy years
ago (see the work of Pauli, Ref.~\cite{Pa1943}),
and have been the subject of recent 
investigations~\cite{BeBo1998,BeDu1999,BeBoMe1999,BeBrJo2002,%
Mo2002i,Mo2002ii,Mo2002iii,Mo2003npb}
because under rather general assumptions, 
they have real energy eigenvalues, or resonance eigenvalues 
which come in complex-conjugate pairs.

This can be seen as follows and holds even though the 
the scalar produce $\left< \psi_1 | \calP | \psi_2 \right>$
is not positive-semidefinite. 
First we recall that because the spectrum of a Hermitian
adjoint operator consists of the complex conjugate eigenvalues, 
there will be an eigenvector $\phi(\vec r)$ with eigenvalue 
$E^*$ provided there exists an eigenvector $\psi(\vec r)$ with 
eigenvalue $E$,
\begin{equation}
H_5(\vec r) \, \psi(\vec r) = E \, \psi(\vec r) \,, 
\qquad
H_5^\plus(\vec r) \, \phi(\vec r) = E^* \, \phi(\vec r) \,.
\end{equation}
The transformation $\vec r \to - \vec r$ and
the insertion of the parity matrix $P = \gamma^0$ leads to 
\begin{align}
H_5^+(-\vec r) \, \phi(-\vec r) =& \; E^* \, \phi(-\vec r) \,,
\nonumber\\[0.77ex]
P H_5^+(-\vec r) P^{-1} \, \left( P \phi(-\vec r) \right) =& \;
E^* \, P \phi(-\vec r) \,.
\end{align}
By assumption, $P H_5^+(-\vec r) P^{-1} = H_5(\vec r)$ and thus
\begin{align}
H_5(\vec r) \, P \phi(-\vec r) =& \; E^* \, P \phi(-\vec r) \,,
\nonumber\\[0.77ex]
H_5(\vec r) \, \widetilde\psi(\vec r) =& \; E^* \widetilde\psi(\vec r) \,,
\qquad
\widetilde\psi(\vec r) = P \phi(-\vec r) \,.
\end{align}
This implies that $\widetilde\psi(\vec r) = P \phi(-\vec r)$ is an eigenvector
with eigenvalue $E^*$. The
eigenvalues of $H_5$ thus come in complex-conjugate pairs.
Furthermore, an additional
quasi-pseudo-Hermitian property can be established,
\begin{equation}
H_5(\vec r) = - \rho \, H^\plus_5(\vec r) \, 
\rho^{-1} \,,
\qquad
\rho = \gamma^0 \, \gamma^5 \,,
\qquad
\rho = - \rho^{-1} \,.
\end{equation}
Going through the above derivation again, we easily find that 
if $E$ and $E^*$ are energy eigenvalues, then 
$-E$ and $-E^*$ will also be energy eigenvalues.
This configuration of eigenvalues can be characterized 
as a ``St.~Andrew's cross configuration'' in the complex plane
(see also Ref.~\cite{Je2012imag}).

For a number of operators such as the $\calP \calT$ 
symmetric variants of odd anharmonic 
oscillators~\cite{BeWe2001,JeSuZJ2009prl,JeSuZJ2010},
the eigenvalues are all real. However, for the 
tachyonic Dirac Hamiltonian, we shall need the full 
spectrum which consists of both real as well as complex resonance eigenvalues.
In order to characterize the spectrum of resonances, 
it is instructive to show that the resonance energies of the 
tachyonic Dirac Hamiltonian are real in the second-order 
formalism. A solution of the tachyonic equation
\begin{equation}
\left( \ii \, \gamma^\mu \, \frac{\partial}{\partial x^\mu} 
- \gamma^5 \, m \right) \, \psi(x) = 0  \,,
\qquad
\psi(x) = U \, \ee^{-\ii \, k \cdot x} \,,
\end{equation}
with $k = (E, \vec k)$, must also fulfill the 
Hamiltonian eigenvalue equation 
\begin{equation}
\left( \vec \alpha \cdot \vec k + \beta \, \gamma^5 \, m \right) \, U = 
E \, U \,,
\end{equation}
and for 
$\psi(x) = V \, \ee^{\ii \, k \cdot x}$ we have
\begin{equation}
\left( -\vec \alpha \cdot \vec k + \beta \, \gamma^5 \, m \right) \, V = 
-E \, V \,.
\end{equation}
Squaring these equations, we thus have
\begin{equation}
\left( \vec \alpha \cdot \vec k + \beta \, \gamma^5 \, m \right)^2 \, U = 
\left( \vec k^{\,2} - m^2 \right) \, U = E^2 \, U \,,
\end{equation}
and 
\begin{equation}
\left( -\vec \alpha \cdot \vec k + \beta \, \gamma^5 \, m \right)^2 \, V = 
\left( \vec k^{\,2} - m^2 \right) \, V = E^2 \, V \,.
\end{equation}
We here observe that the matrix $\beta \, \gamma^5$ is a 
representation of the imaginary unit, in view of 
the identity $(\beta \, \gamma^5)^2 = - \mathbbm{1}_{4 \times 4}$.
Connections to other representations of the 
imaginary unit within superluminal spin-$1/2$ Hamiltonians
are discussed in~\ref{appa}.
The superluminal energy-momentum dispersion relation
\begin{equation}
\label{dispersion2}
E^2 = \vec k^{\,2} - m^2
\end{equation}
(with real $E$) is thus fulfilled. 
Resonance energies $E$ of the tachyonic Dirac Hamiltonian 
thus have to be either completely real (vanishing imaginary part) or
completely imaginary (vanishing real part), so that $E^2$ is real.

%
%
\subsection{Need for the Inclusion of Resonances and Anti--Resonances}

One might wonder about the nature of plane-wave
solutions with $|\vec k| < m$, which according to 
$E = \sqrt{\vec k^2 - m^2}$ would have imaginary
energies. Denoting by $u > 1$ the velocity of the 
particle (expressed in units of the speed of light $c$), 
we have for tachyonic particles, 
according to Refs.~\cite{BiDeSu1962,Fe1967,ArSu1968,DhSu1968,BiSu1969},
the relations
\begin{equation}
\label{tachdisp}
E = \frac{m}{\sqrt{u^2 - 1}} \,,
\qquad
|\vec k| = |\vec p|  = \frac{m \, u}{\sqrt{u^2 - 1}} > m \,.
\end{equation}
where $u$ is the particle velocity in units of the 
velocity of light, $c$.
Superluminal Lorentz transformations have been discussed 
in Refs.~\cite{BiDeSu1962,SuSh1986} 
(see also Fig.~\ref{fig1}); it is impossible to 
`stop' a superluminal particle, and the `rest frame' of a 
superluminal particle is that of infinite velocity~\cite{Fe1967,BaSh1974}.
We conclude that since $|\vec k| > m$,
the problematic imaginary energies which are 
encountered for $|\vec k| < m$ are excluded 
by the tachyonic dispersion relation.
This leads to a restriction $|\vec k| > m$
for the permissible values in constructing stable 
tachyonic wave packets (``Fourier transformations'')
from the fundamental plane-wave solutions.
In a distant analogy to the ``positive definite''
functions which are the subject of Bochner's theorem
(Theorem IX.9 of Ref.~\cite{ReSi1978vol2}),
the wave packets of tachyonic particles have to 
be ``positive definite'' in the sense that they 
are naturally constructed using plane waves with 
$|\vec k| > m$ in view of the tachyonic dispersion relation. 
(A ``positive definite'' function 
has a positive Fourier transform, or more precisely, 
is characterized by a Fourier transform which constitutes
a positive Borel measure.)

In Refs.~\cite{BiDeSu1962,Fe1967,ArSu1968,DhSu1968},
considerable effort has thus been invested into 
the elimination of the states with $\vec k^2 < m^2$ 
from the scalar tachyonic theory, because it was realized that 
the resonances correspond to evanescent waves.
However, we here find that these states qualify themselves
as resonances of the tachyonic Dirac Hamiltonian.

We define the width $\gamma$ of a resonance of the tachyonic 
Dirac Hamiltonian as follows,
\begin{align}
\label{EEEGAMMA}
E =& \; \pm \sqrt{\vec k^{\,2} - m^2 - \ii \, \epsilon} = 
\mp \, \ii \, \frac{\Gamma}{2} \,,
\qquad
\Gamma = 2 \,\sqrt{m^2 - \vec k^{\,2}}  \,, 
\qquad
| \vec k | < m \,.
\end{align}
The wave functions describing the resonances are as follows,
\begin{subequations}
\label{RR}
\begin{align}
R_+(\vec k) = & \;
\left( \begin{array}{c}
\dfrac{m+\tfrac{\ii}{2} \Gamma +|\vec k|}{\sqrt{2} \, 
\sqrt{\vec k^{\,2} + m^2 + \tfrac14 \, \Gamma^2}} \; a_+(\vec k) \\[0.77ex]
\dfrac{m-\tfrac{\ii}{2} \Gamma -|\vec k|}{\sqrt{2} \, 
\sqrt{\vec k^{\,2} + m^2 + \tfrac14 \, \Gamma^2}} \; a_+(\vec k) \\
\end{array} \right) \,,
\\[0.77ex]
R_-(\vec k) = & \;
\left( \begin{array}{c}
\dfrac{m-\tfrac{\ii}{2} \Gamma -|\vec k|}{\sqrt{2} \, 
\sqrt{\vec k^{\,2} + m^2 + \tfrac14 \, \Gamma^2}} \; a_-(\vec k) \\[0.77ex]
\dfrac{-m-\tfrac{\ii}{2} \Gamma-|\vec k|}{\sqrt{2} \, 
\sqrt{\vec k^{\,2} + m^2 + \tfrac14 \, \Gamma^2}} \; a_-(\vec k) \\
\end{array} \right) \,,
\\[0.77ex]
E =& \; -\tfrac{\ii}{2}\, \Gamma = -\tfrac{\ii}{2} \, \sqrt{m - \vec k^{2}} \,,
\qquad
\vec k^{\,2} < m^2 \,.
\end{align}
\end{subequations}
The antiresonance eigenstates are
\begin{subequations}
\label{SS}
\begin{align}
S_+(\vec k) = & \;
\left( \begin{array}{c}
\dfrac{-m-\tfrac{\ii}{2} \Gamma+|\vec k|}{\sqrt{2} \, 
\sqrt{\vec k^{\,2} + m^2 + \tfrac14 \, \Gamma^2}} \; a_+(\vec k) \\[0.77ex]
\dfrac{-m+\tfrac{\ii}{2} \Gamma-|\vec k|}{\sqrt{2} \, 
\sqrt{\vec k^{\,2} + m^2 + \tfrac14 \, \Gamma^2}} \; a_+(\vec k) \\
\end{array} \right) \,,
\\[0.77ex]
S_-(\vec k) = & \;
\left( \begin{array}{c}
\dfrac{-m+\tfrac{\ii}{2}\Gamma -|\vec k|}{\sqrt{2} \, 
\sqrt{\vec k^{\,2} + m^2 + \tfrac14 \, \Gamma^2}} \; a_-(\vec k) \\[0.77ex]
\dfrac{m+\tfrac{\ii}{2}\Gamma -|\vec k|}{\sqrt{2} \, 
\sqrt{\vec k^{\,2} + m^2 + \tfrac14 \, \Gamma^2}} \; a_-(\vec k) \\
\end{array} \right) \,,
\\[0.77ex]
E =& \; \tfrac{\ii}{2}\, \Gamma = \tfrac{\ii}{2} \, \sqrt{m - \vec k^{2}} \,,
\qquad \vec k^{\,2} < m^2 \,.
\end{align}
\end{subequations}
These states are normalized, i.e., $R^\plus_+(\vec k) \, R_+(\vec k) =
R^\plus_-(\vec k) \, R_-(\vec k) = S^\plus_+(\vec k) \, S_+(\vec k) =
S^\plus_-(\vec k) \, S_-(\vec k) = 1$.
For $|\vec k| = m$, we have $E \to 0$, and the eigenstates read as
\begin{subequations}
\label{toUU}
\begin{align}
U_+(\vec k) \to & \;
R_+(\vec k) \to 
\left( \begin{array}{c}
a_+(\vec k) \\[0.77ex] 0 \\
\end{array} \right) \,,
\qquad |\vec k| \to m \,,
\\[0.77ex]
U_-(\vec k) \to & \;
R_-(\vec k) \to 
\left( \begin{array}{c}
0 \\[0.77ex] -a_-(\vec k) \\
\end{array} \right) \,,
\qquad |\vec k| \to m \,.
\end{align}
\end{subequations}
The negative-energy spinors tend to the following values,
\begin{subequations}
\label{toVV}
\begin{align}
V_+(\vec k) \to & \;
S_+(\vec k) \to 
\left( \begin{array}{c}
0 \\[0.77ex] -a_+(\vec k) \\
\end{array} \right) \,,
\qquad |\vec k| \to m \,,
\\[0.77ex]
V_-(\vec k) \to & \;
S_-(\vec k) \to 
\left( \begin{array}{c}
-a_-(\vec k) \\[0.77ex] 0 \\ 
\end{array} \right) \,,
\qquad |\vec k| \to m \,.
\end{align}
\end{subequations}
According to Eq.~\eqref{tachdisp}, these spinors, which correspond to 
a tachyon of infinite speed, correspond to the solutions which 
describe a tardyon spinor at rest. This point has also been 
made in Ref.~\cite{BaSh1974}.

At time $t=0$, the above vectors are eigenstates 
of the Hamiltonian,
\begin{subequations}
\label{complete}
\begin{align}
\psi_{1,\vec k}(x) =& \; 
U_\pm(\vec k) \, \ee^{\ii \vec k \cdot \vec r} \,,
\qquad |\vec k| \geq m \,,
\\[0.77ex]
\psi_{2,\vec k}(x) =& \; 
V_\pm(-\vec k) \, \ee^{\ii \vec k \cdot \vec r} \,,
\qquad |\vec k| \geq m \,,
\\[0.77ex]
\psi_{3,\vec k}(x) =& \; 
R_\pm(\vec k) \, \ee^{\ii \vec k \cdot \vec r} \,,
\qquad |\vec k| < m \,,
\\[0.77ex]
\psi_{4,\vec k}(x) =& \; 
S_\pm(-\vec k) \, \ee^{\ii \vec k \cdot \vec r} \,,
\qquad |\vec k| < m \,,
\end{align}
\end{subequations}
with $\vec k \in \mathbbm{R}^3$.
These form a complete set of eigenstates of the 
tachyonic Dirac Hamiltonian and allow us to 
solve the initial value problem,
\begin{equation}
\ii \, \partial_t \psi(t, \vec r) = H_5 \, \psi(t, \vec r) \,,
\qquad
\psi(0, \vec r) = \Psi(\vec r) \,,
\end{equation}
where the initial wave-function configuration is given by~$\Psi(\vec r)$.
We should indicate a little caveat, here:
The $U_\pm(\vec k)$ are mutually orthogonal 
(positive energy), even for the same $\vec k$, 
and the $V_\pm(\vec k)$ (solutions for negative energy)
have the same property [see Eq.~\eqref{UUVV} below].
The initial value problem is thus 
solved on the level of the relativistic quantum theory
for both positive-energy as well as negative-energy 
wave packets. The only remaining problem is the non-unitary character of the 
time evolution generated by the tachyonic Dirac Hamiltonian 
through the resonances, and antiresonances. Such a non-unitary
time-evolution in the presence of resonances is discussed in 
detail in Ref.~\cite{JeSuLuZJ2008}
for the case of the odd anharmonic oscillator.
We should clarify that 
the resonance solutions $R_\pm(\vec k)$ and $S_\pm(\vec k)$ 
are not resonances in the sense of a virtual state in a cross
section (which ``decays'' into a final state), 
but constitute complex-valued resonance energies of the Hamiltonian matrix.
They describe the evanescent subluminal components of a wave
packet which is genuinely superluminal.
According to the $\ii \epsilon$ prescription from Eq.~\eqref{EEEGAMMA}, 
resonance solutions constitute the 
analytic continuation of positive-energy solutions 
for $|\vec k| < m$. They  are exponentially
damped for propagation into the future.
By contrast, anti-resonance solutions
describe the continuation of negative-energy solutions 
for $|\vec k| < m$,
these are damped for propagation into the past.
This prescription ensures that an inconsistent exponential 
growth of the wave packets cannot occur~\cite{JeWu2012}.

Our complete set of functions (not restricted to $|\vec k| < m$)
allows us to 
represent localized wave packets in terms of a Fourier decomposition.
One example of a localized wave function in a scalar theory is
\begin{equation}
\delta^3(\vec r) = \int \frac{\dd^3 k}{(2\pi)^3} \, 
\ee^{\ii \vec k \cdot \vec r} \,.
\end{equation}
It has the property
\begin{equation}
\int \dd^3 r' \, \delta^3(\vec r') \, \delta^3(\vec r + \vec r') =
\delta^3(\vec r) \,.
\end{equation}
Likewise the function
\begin{align}
\Psi(\vec r) = & \;
\int_{|\vec k|<m} \frac{\dd^3 k}{(2\pi)^3} \, 
R_{+}(\vec k)\ee^{\ii \vec k \cdot \vec r} 
+ \int_{|\vec k| \geq m} \frac{\dd^3 k}{(2\pi)^3} \, 
U_{+}(\vec k)\ee^{\ii \vec k \cdot \vec r} 
\end{align}
has the property
\begin{align}
& \int \dd^3 r' \, \Psi^\plus(\vec r') \, \Psi(\vec r + \vec r') =
\int \frac{\dd^3 k}{(2\pi)^3} \,
\left| \widetilde \Psi(\vec k) \right|^2 \, 
\ee^{\ii \vec k \cdot \vec r} 
= \int \frac{\dd^3 k}{(2\pi)^3} \,
\ee^{\ii \vec k \cdot \vec r} = 
\delta^3(\vec r) \,.
\end{align}
where $\widetilde \Psi(\vec k)$ is the Fourier transform of $\Psi(\vec r)$.

Three remarks are in order.
{\it (i)}~In an idealized position measurement using a detector,
according to the wave function collapse postulate of quantum theory,
the wave function is proportional to a Dirac-$\delta$ after the measurement.
It is important to be able to localize a tachyonic 
particle in order to measure it. According to the Heisenberg principle,
the position of quantum particle cannot be measured significantly 
better than its Compton wavelength, but this finding does not imply that 
idealized position measurements are meaningless.
In our formalism (see also Ref.~\cite{JeWu2012}), 
we prefer to retain localizability of a particle 
rather than prevent the emergence of the small region of non-unitary time evolution 
in the sector $|\vec k| < m$.
{\it (ii)}~It would be somewhat unnatural to exclude the resonances and 
antiresonances of the tachyonic Dirac Hamiltonian from the 
discussion {\em a fortiori}.
Canonically, resonances and antiresonances play an important 
role in the spectrum of quantum Hamiltonians and should receive a physical
interpretation~\cite{ReSi1978vol3}.
In our case, we propose an interpretation 
in terms of the evanescent waves that describe tardyonic components of 
a genuinely tachyonic wave packet.
{\it (iii)}~For completeness and for general interest, we note that 
the Hamiltonian $H' = \vec \alpha \cdot \vec p + \ii \, \beta \, m$, 
where $\ii$ is the imaginary unit and $m$ is a real mass term, is 
connected with $H_5$ by a unitary transformation, which 
is constructed in~\ref{appa}. 

With regard to remark~(i) and
Refs.~\cite{BiDeSu1962,Fe1967,ArSu1968,DhSu1968,BiSu1969}, we observe that full
unitarity cannot be preserved anyway in a tachyonic theory at the one-loop
level, as observed in Ref.~\cite{Bo1970}. 
Although the unitarity violation at the one-loop level could be 
interpreted as an argument against
the consistency of tachyonic field theories,
we here prefer to countenance a small unitary violation~\cite{JeWu2012}
due to the evanescent waves than to sacrifice the localizability of a
wave packet. Likewise, in Ref.~\cite{XiJi1987}, it was
argued that tachyonic particles cannot be described by a unitary representation
of the Lorentz group. In general, we have to sacrifice either unitarity or
localizability in we wish to incorporate tachyons into field theory.  As
observed above and explained in Ref.~\cite{JeWu2012}, the slight violation of
unitarity is restricted to a relatively small kinematic region.  Upon the
inclusion of the resonances, we observe that one can write field commutators
like those encountered in Eq.~(1.4) of Ref.~\cite{DhSu1968} in terms of the
full Dirac-$\delta$ function instead of the ``filtered'' Dirac-$\delta$
function given in Eq.~(1.5) of Ref.~\cite{DhSu1968}, which is not localized.
This is explained in detail in Ref.~\cite{JeWu2012}.

With regard to remark~(iii), we note that the field theory based on the
Hamiltonian $H'$ was analyzed in Ref.~\cite{BaSh1974}; this analysis has
recently been enhanced in Ref.~\cite{Je2012imag}.

%
%
\subsection{Orthogonality Properties of the Solutions}
\label{ortho_prop}

Finally, a discussion of the orthogonality properties
of the solutions of the tachyonic Dirac equation is in order.
Let us suppose that $\psi_2$ is 
an eigenvector of a pseudo-Hermitian Hamiltonian $H_5$ with 
eigenvalues $E_2$ and $\psi_1$ is an 
eigenvector of $H_5$ with
eigenvalues $E_1$. Then,
\begin{align}
\label{scalar}
\left< \psi_1 | \calP | \psi_2 \right> \equiv & \;
\int \dd^3 x \, \psi^\plus_1(\vec x) \gamma^0 \psi_2(-\vec x) \,,
\nonumber\\[2ex]
E_2 \left< \psi_1 | \calP | \psi_2 \right> =& \;
\left< \psi_1 | \calP E_2 | \psi_2 \right> =
\left< \psi_1 | \calP H_5 | \psi_2 \right> 
\nonumber\\[2ex]
=& \; \left< \psi_1 | \calP H_5 \calP^{-1} \calP | \psi_2 \right> =
\left< \psi_1 | H_5^\plus \calP | \psi_2 \right> 
\nonumber\\[2ex]
=& \; \left< H_5 \psi_1 | \calP | \psi_2 \right> =
E_1^* \left< \psi_1 | \calP | \psi_2 \right> \,,
\end{align}
i.e., eigenvectors for different energy are
orthogonal with respect to the $\calP$-scalar product.
Our plane-wave solution form a complete set of eigenstates of the
Hamiltonian $H_5$, with the properties
\begin{subequations}
\label{ortho}
\begin{align}
H_5 \, U_\pm(\vec k) \, \ee^{\ii \vec k \cdot \vec r} = & \;
E \, U_\pm(\vec k) \, \ee^{\ii \vec k \cdot \vec r}\,, \\[2ex]
H_5 \, V_\pm(-\vec k) \, \ee^{\ii \vec k \cdot \vec r} = & \;
-E \, V_\pm(-\vec k) \, \ee^{\ii \vec k \cdot \vec r} \,, \\[2ex]
\label{UUVV}
U^\plus_\alpha(\vec k) \, U_\beta(\vec k) =& \; \delta_{\alpha\beta} \,,
\qquad
V^\plus_\alpha(-\vec k) \, V_\beta(-\vec k) =
\delta_{\alpha\beta} \,,\\[2ex]
U^\plus_\alpha(\vec k) \gamma^0 V_\beta(+\vec k) =& \; 0 \,, \qquad
\alpha,\beta = \pm \,,
\end{align}
\end{subequations}
and $E = \sqrt{\vec k^2 - m^2}$.
The relation $U^\plus_\alpha(\vec k) \gamma^0 V_\beta(+\vec k) = 0$
is equivalent to 
\begin{align}
& \left< U_\pm(\vec k) \, \ee^{\ii \vec k \cdot \vec r} \left| 
\calP \right| V_\pm(\vec k) \, \ee^{-\ii \vec k \cdot \vec r} \right> 
= \left< U_\pm(\vec k) \, \ee^{\ii \vec k \cdot \vec r} \left| 
\gamma^0 \right| V_\pm(+\vec k) \, \ee^{\ii \vec k \cdot \vec r} \right> = 0 \,,
\end{align}
which is equivalent to the orthogonality of the eigenvectors 
with different energies $E$ and $-E$ under the scalar product defined in 
Eq.~\eqref{scalar}.
Within the space of positive-energy and negative-energy 
solutions, the time evolution is unitary with respect
to the ordinary scalar product defined in Eq.~\eqref{UUVV}.
Furthermore, we observe that while 
in general $U^\plus_\alpha(\vec k) V_\beta(-\vec k) \neq 0$,
the four solutions $U_\pm(\vec k) \, \ee^{\ii \vec k \cdot \vec r}$ and
$V_\pm(-\vec k) \, \ee^{\ii \vec k \cdot \vec r}$
form a linearly independent set for given $\vec k$.
However, as explained in~\ref{appb}, the $\calP$ norm of 
the plane-wave states with a nonvanishing $\vec k$ actually vanishes,
and the $\calP$ scalar product therefore is not 
positive definite. Therefore, as explained in detail 
in Ref.~\cite{BeBrReRe2004}, it would be 
desirable to work with a different scalar product, 
obtained by prepending the $\calP$ operator with a 
second operator, called $\calC$ in Ref.~\cite{BeBrReRe2004}
(while being manifestly different from charge conjugation,
see~\ref{appb}), so that the ensuing redefined scalar product is 
positive-definite. 

The construction of a positive-definite scalar product, 
conserved under the pseudo-Hermitian time evolution,
is left as an open problem for future investigations. 
However, one important observation can be made at the current 
stage: namely, the time evolution induced by the 
pseudo-Hermitian Hamiltonian is separately unitary for 
wave packets constructed from exclusively positive-energy 
and negative-energy wave packets, in view of the property~\eqref{UUVV}
and the reality of the eigenvalues. 
This pseudo-unitarity is sufficient if one assumes that the 
Hamiltonian naturally breaks up into a positive-energy 
and negative-energy Hamiltonian, according to the 
interpretation of positive-energy solutions describing 
particles and negative-energy solutions describing 
antiparticles~\cite{JeWu2012}.
Indeed, for a Hermitian Hamiltonian, the scalar product
\begin{subequations}
\begin{equation}
\langle \psi_1(t) | \psi_2(t) \rangle  = 
\int \dd^3 x \, \psi^+(\vec r, t) \, \psi(\vec r, t)
\end{equation}
is conserved, as can be seen by differentiation with respect
to time. For a pseudo--Hermitian Hamiltonian which fulfills
Eq.~\eqref{pseudoH}, time evolution 
is unitary with respect to the scalar product
$\langle \psi_1(t) | \calP | \psi_2(t) \rangle$
(see Ref.~\cite{Pa1943}).
This is because
\begin{align}
& \ii \, \partial_t \langle \psi_1(t) | \calP | \psi_2(t) \rangle = 
\int \dd^3 x \, \left(-\ii \partial_t \psi_1(t, \vec r) \right)^+ \, 
\calP \psi_2(t, \vec r)
+ \int \dd^3 x \, \psi_1^+(t, \vec r) \, \calP \; \ii \partial_t \psi_2(t, \vec r)
\nonumber\\[0.77ex]
& \; = \int \dd^3 x \, \left(-H_5(\vec r) \, \psi_1(t, \vec r) \right)^+ \, 
\calP \psi_2(t, \vec r)
+ \int \dd^3 x \, \psi_1^+(t, \vec r) \, \calP \, H_5(\vec r) \psi_2(t, \vec r)
\nonumber\\[0.77ex]
& \; = -\langle \psi_1(t) | H^+_5(\vec r) \calP | \psi_2 \rangle
+ \langle \psi_1 | \calP \, H_5(\vec r) | \psi_2 \rangle = 0 \,,
\end{align}
where we use the property 
$H^+_5(\vec r) \calP  =  \calP \, H_5(\vec r)$.
Thus, $\langle \psi_1(t) | \calP | \psi_2(t) \rangle$
is conserved under pseudo-Hermitian time evolution.
However, for wave packets constructed from plane 
waves that fulfill the superluminal dispersion relation~\eqref{tachdisp},
we can actually show a stronger statement because the 
energy eigenvalues $E_{\vec k} = \sqrt{\vec k^2 - m^2}$ with 
$|\vec k| > m$ are purely real. 
Let us study two {\em positive-energy} 
normalized wave packets of the form ($j=1,2$),
\begin{align}
& \qquad \psi_j(t, \vec r) = \int \frac{\dd^3 k}{(2\pi)^3} \, 
\left( w_{j,+}(\vec k) \, U_+(\vec k) +
w_{j,-}(\vec k) \, U_-(\vec k) \right) \,
\ee^{-\ii E_{\vec k} t + \ii \vec k \vec r}  \,,
\nonumber\\
& \qquad
\int \dd^3 k \, \left( |w_{j,+}(\vec k)|^2 + |w_{j,-}(\vec k)|^2 \right) = 1 \,,
\end{align}
where the $U_\pm(\vec k)$ are given in Eq.~\eqref{UU}.
Then,
\begin{align}
\langle  \psi_1(t, \vec r) | \psi_2(t, \vec r) \rangle =& \;
\int \dd^3 x \int \frac{\dd^3 k'}{(2\pi)^3} \,
\int \frac{\dd^3 k''}{(2\pi)^3} \,
\left( w^*_{1,+}(\vec k') \, U^\plus_+(\vec k') +
w^*_{1,-}(\vec k') \, U^\plus_-(\vec k') \right) \,
\nonumber\\ & \; \times
\left( w_{2,+}(\vec k'') U_+(\vec k'') +
w_{2,-}(\vec k'') U_-(\vec k'') \right) 
\ee^{-\ii (E_{\vec k''} - E_{\vec k'}) t +
\ii (\vec k'' - \vec k') \cdot \vec r}  
\nonumber\\
=& \; 
\int \frac{\dd^3 k'}{(2\pi)^3} \,
\int \dd^3 k'' \,
\left( 
w^*_{1,+}(\vec k') \, w_{2,+}(\vec k'') \, U^\plus_+(\vec k') \, U_+(\vec k'') 
\right.
\nonumber\\
& \; \left. 
+ w^*_{1,-}(\vec k') \, w_{2,-}(\vec k'') \, U^\plus_-(\vec k') \, U_-(\vec k'') \right)
\ee^{-\ii (E_{\vec k''} - E_{\vec k'}) t } 
\delta^3(\vec k'' - \vec k') 
\nonumber\\
=& \;
\int \frac{\dd^3 k}{(2\pi)^3} \,
\left( 
w^*_{1,+}(\vec k) \, w_{2,+}(\vec k) +
w^*_{1,-}(\vec k) \, w_{2,-}(\vec k) \right) 
\nonumber\\
=& \; \langle  \psi_1(0, \vec r) | \psi_2(0, \vec r) \rangle \,.
\end{align}
\end{subequations}
Thus, unitary time evolution holds provided the 
energies $E_{\vec k} = \sqrt{\vec k^2 - m^2}$ 
fulfill the dispersion relation given in Eq.~\eqref{tachdisp},
which implies that they are manifestly real.
Within the manifolds of positive-energy and 
negative-energy wave packets, the $\calC$ operator
in the sense of~\ref{appb} thus can be taken to be equal 
to the parity operator $\calP$, in accordance with the 
findings of Ref.~\cite{BeBrChWa2005CP}.

%
%
\section{Conclusions}
\label{conclu}

In the current article, the tachyonic Dirac equation
$\left( \ii \gamma^\mu \partial_\mu - \gamma^5 m \right) \, \psi = 0 $
is analyzed. We show that it is invariant under time reversal ($\calT$,
Sec.~\ref{SecFifthT}), and invariant under charge conjugation and parity
($\calC\calP$).  Our formulation of the
time reversal operation involves the Dirac adjoint (see Sec.~4.4.2 of
Ref.~\cite{HeLect}).  We generalize the treatment of the
$\calC$, $\calP$ and $\calT$ transformations to the tachyonic Dirac
equation in Secs.~\ref{SecFifthC},~\ref{SecFifthP} and~\ref{SecFifthT}.
These are the desired symmetry properties for the description of 
neutrinos.
The solutions of the tachyonic equation
approximate the chiral solutions of the massless
Dirac equation in the limit $m \to 0$
(see Sec.~\ref{solutions}). 
In writing the solutions, one is naturally 
led to the use of a basis of chiral eigenfunctions,
aligned with the propagation four-momentum 
$k = (E, \vec k)$ of the particle.
The tachyonic Hamiltonian given in Eq.~\eqref{fifthH},
$H_5 = \vec \alpha \cdot \vec p + \beta \, \gamma^5 \, m$,
is shown to be pseudo--Hermitian.
The concept of pseudo--Hermiticity is related to an 
article of Pauli~\cite{Pa1943} who established Hamiltonians 
with the given property as viable alternatives to Hermitian 
Hamiltonians
(see also Refs.~\cite{BeBo1998,BeDu1999,BeBoMe1999,BeBrJo2002}). 
In particular, pseudo-Hermiticity
has been indispensable in the recent generalization of the 
so-called Bender--Wu formulas~\cite{BeWu1969,BeWu1971,BeWu1973}
to odd anharmonic oscillators~\cite{JeSuZJ2009prl,JeSuZJ2010}.

In the tachyonic Dirac Hamiltonian 
$H_5 = \vec \alpha \cdot \vec p + \beta \, \gamma^5 m$,
the matrix $\ii_4 = \beta \, \gamma^5$ is a
representation of the imaginary unit in four dimensions,
\begin{equation}
\ii_4 = \beta \, \gamma^5 =
\left( \begin{array}{cc} 0 & \mathbbm{1}_{2\times 2} \\ 
-\mathbbm{1}_{2\times 2} & 0 \end{array} \right)\,,
\qquad
\left( \ii_4 \right)^2 = - \mathbbm{1}_{4\times 4}\,,
\end{equation}
where $\mathbbm{1}_{2\times 2}$ denotes the $2 \times 2$ unit matrix,
and $\mathbbm{1}_{4\times 4}$ denotes the $4 \times 4$ unit matrix.
Indeed, it was Dirac who argued~\cite{Di1928a,Di1928b} that the 
linearization of the Klein--Gordon equation required 
the introduction of $4 \times 4$ matrices; the most straightforward
implementation of the imaginary unit in four-dimensional 
space is due to the matrix $\ii_4 = \beta \, \gamma^5$
because it generalizes the two-dimensional representation 
of the complex imaginary unit. Indeed, it is well known that 
complex multiplication can be written as matrix multiplication
within the group of matrices of the form
\begin{equation}
z = |z| \, 
\left( \begin{array}{cc} \cos\theta & \sin\theta \\ 
-\sin\theta & \cos\theta \end{array} \right) =
\left( \begin{array}{cc} {\rm Re} \, z &  {\rm Im} \, z \\
-{\rm Im} \, z & {\rm Re} \, z \end{array} \right)\,,
\end{equation}
where $z = |z| \exp(\ii \, \theta)$ 
is a complex number and the imaginary unit has 
$|z| = 1$ and $\theta = \pi/2$.
Division algebras (``matrix multiplication algebras'')
without zero divisors only exist in
1, 2, 4 and 8 dimensions~\cite{BoMi1958,Ke1958}.

We have discussed the description of a tachyonic wave packet using 
a pseudo-Hermitian Hamiltonian whose plane-wave solutions fulfill the 
tachyonic dispersion relation. The Hamiltonians 
$H_5 = \vec \alpha \cdot \vec p + \beta \, \gamma^5 \, m$,
or alternatively $H' = \vec \alpha \cdot \vec p + \ii \, \beta \, m$,
have pseudo-Hermitian as well as 
quasi-pseudo-Hermitian properties, as discussed in Sec.~\ref{solutions}.
The description leads to a 
unitary time evolution within wave packets that  
are composed exclusively of positive-energy and 
negative-energy waves. The scalar product employed in the 
calculation of the unitary time evolution is the ordinary 
scalar product, and it is positive-definite for 
positive-energy and negative-energy wave packets. 
Positive-energy solutions are 
commonly interpreted as those characterizing particles,
and negative-energy solution characterize antiparticles.
It has recently been shown~\cite{JeWu2012} that for neutrinos,
the field-theoretical generalization of the derivation presented 
here leads to a natural suppression of right-handed neutrinos
and left-handed anti-neutrinos, 
due to an indefinite Hilbert-space norm of the 
``wrong-helicity'' states which is induced by helicity-dependent 
anticommutators, as explained in Ref.~\cite{JeWu2012}.
On the level of quantum mechanics, 
the construction of a conserved (under time evolution) 
scalar product applicable to both positive-energy 
as well as negative-energy wave packets is left as an open problem.
However, we stress that due to the $\ii \epsilon$ prescription 
as employed in Ref.~\cite{JeWu2012} in the tachyonic propagator 
denominator, positive-energy and negative-energy 
solutions propagate independently. 

Inconclusive indications for a possibly superluminal nature of the 
neutrino come from low-energy beta decay 
experiments~\cite{RoEtAl1991,AsEtAl1994,StDe1995,AsEtAl1996,%
WeEtAl1999,LoEtAl1999,BeEtAl2008},
from an early arrival of neutrinos from the 1987 supernova~\cite{DaEtAl1987},
from the MINOS laboratory-based time-of-flight experiment
(as reported in Ref.~\cite{AdEtAl2007}),
as well as from an earlier FERMILAB experiment~\cite{KaEtAl1979}.
The OPERA experiment~\cite{OPERA2011v2} is currently 
being reexamined. While none of these experiments provides conclusive 
evidence for the superluminal nature of the neutrino, 
it is interesting to observe that the experimentally
determined best estimate of the neutrino propagation
velocity has been measured as superluminal in 
Refs.~\cite{KaEtAl1979,AdEtAl2007,OPERA2011v2},
and the best estimates for the neutrino mass squares have
been determined to be negative in 
Refs.~\cite{RoEtAl1991,AsEtAl1994,StDe1995,AsEtAl1996,%
WeEtAl1999,LoEtAl1999,BeEtAl2008},
with a tendency toward smaller absolute values for the 
mass squares at lower energies
(for a summary see also~\cite{LABneutrino}).
Conceivable energy-dependent neutrino mass running is discussed in 
Refs.~\cite{LiWa2011,Je2012cejp}.

\appendix 

%
%
\section{Unitary Transformations of the Superluminal Hamiltonian}
\label{appa}

In Refs.~\cite{BaSh1974,Je2012imag}, the Hamiltonian
\begin{equation}
H' = \vec \alpha \cdot \vec p + \ii \, \beta \, m \,,
\end{equation}
is investigated, where $\ii$ is the imaginary unit and $m$ is a real
mass term. This Hamiltonian is related to the 
tachyonic Dirac Hamiltonian as follows. We consider the 
unitary transformation
\begin{align}
U =& \; \frac{1}{\sqrt{2}} \,
\left[ \left( \frac12 + \frac{\ii}{2} \right) \, 
\left( 1 - \gamma^0 \, \gamma^5 \right) +
\left( \frac12 - \frac{\ii}{2} \right) \, 
\left( \gamma^0 + \gamma^5 \right) \right] \, \gamma^1 \,,
\nonumber\\[2ex]
=& \; 
\left( \begin{array}{cccc}
0 & \frac{\ii}{\sqrt{2}} & 0 & \frac{1}{\sqrt{2}} \\
\frac{\ii}{\sqrt{2}} & 0 & \frac{1}{\sqrt{2}} & 0 \\
0 & -\frac{\ii}{\sqrt{2}} & 0 & \frac{1}{\sqrt{2}} \\
-\frac{\ii}{\sqrt{2}} & 0 & \frac{1}{\sqrt{2}} & 0 \\
\end{array} \right)  \,,
\end{align}
which has the property
\begin{equation}
U \beta \, \gamma^5 \, U^{-1} = \ii \, \beta\,,
\qquad
\widetilde{\alpha}^i = U \, \alpha^i \, U^{-1} \,,
\qquad
\left\{ \alpha^i , \, \alpha^j \right\} = 
\left\{ \widetilde\alpha^i , \, \widetilde\alpha^j \right\} = 2 \, \delta^{ij} \,.
\end{equation}
Now,
\begin{equation}
H'' = U \, H_5 \, U^{-1} =
U \, \left( \alpha^i \, p^i + \beta \, \gamma^5 \, m \right) \, U^{-1} = 
\widetilde{\alpha}^i \, p^i + \ii \, \beta m \,.
\end{equation}
Because the $\widetilde{\alpha}^i$ fulfill the same algebraic
relations as the ${\alpha}^i$, the Hamiltonian $H''$ is equivalent to 
the Hamiltonian ($\widetilde{\alpha}^i \to \alpha^i$) 
\begin{equation}
H' = \alpha^i \, p^i + \ii \, \beta \, m \,.
\end{equation}
Using a further unitary transformation,
we may invert the sign of the imaginary mass term,
\begin{equation}
H''' = \gamma^5\, H' \, (\gamma^5)^{-1} = \alpha^i \, p^i - \ii \, \beta \, m \,,
\end{equation}
where $\beta^{-1} = \beta$ is a unitary operator.
Because of the unitary relation of $H'''$, $H''$, $H'$, and $H_5$,
it is not surprising that the field theories induced by the 
Hamiltonians are indeed equivalent, 
as analyzed in detail in Ref.~\cite{Je2012imag}
(with a special emphasis on $H_5$ and $H'$).

%
%
\section{Quantum Mechanics and Scalar Products}
\label{appb}

We attempt to investigate the connection to 
quantum mechanical Hamiltonians as described in 
Refs.~\cite{BeBo1998,BeDu1999,BeBoMe1999,BeWe2001,BeBrJo2002}.
For definiteness, we follow the conventions of 
Ref.~\cite{BeBrReRe2004},
which implies that the operators $\calP$ and $\calT$ 
are restricted to the domain of ordinary 
(nonrelativistic) quantum mechanical operators,
{\em and from now one, 
$\calC$ is redefined to be a pseudo charge-conjugation
in accordance with Ref.~\cite{BeBrReRe2004}}.
Hamiltonians of the following structure are being investigated,
\begin{equation}
\label{HPT}
H = p^2 + x^2 \, \left( \ii \, x\right)^\epsilon =
- \frac{\partial^2}{\partial x^2} + 
x^2 \, \left( \ii \, x\right)^\epsilon \,,
\end{equation}
where $p =-\ii \, \partial/\partial x$ is the momentum operator.
In Refs.~\cite{BeBo1998,BeDu1999,BeBoMe1999,BeWe2001,BeBrJo2002,BeBrReRe2004}, 
the parity operator is defined to be equivalent to 
the transformation $x \to -x$, $p \to -p$, 
while the $\calT$ operator describes the time inversion, which is 
equivalent to $x \to x$, $p \to -p$, and $\ii \to -\ii$.
We note that the transformation $\ii \to -\ii$ is equivalent to the 
calculation of the Hermitian adjoint.
For the class of Hamiltonians given by Eq.~\eqref{HPT},
which we consider exclusively from now on, the
$\calT$ transformation is equivalent to 
the calculation of the Hermitian adjoint, and we have
\begin{equation}
\left[ \calP \, \calT, H \right] = 0 \,,
\qquad
H = \calP \, \calT \, H \, \calT^{-1} \, \calP^{-1} 
= \calP \, H^\plus \, \calP^{-1} \,.
\end{equation}
In the case of unbroken $\calP \cal T$ symmetry, 
the eigenfunctions of the Hamiltonian are also 
eigenfunctions of the $\calP \calT$ operator~\cite{BeBrReRe2004},
\begin{equation}
\label{HPTprop}
H \, \phi_n(x) = E_n \, \phi_n(x) \,,
\qquad
\calP \, \calT \, \phi_n(x) = \phi_n(x) \,.
\end{equation}
Using the Hermiticity of the $\calP$ operator, we can formulate
the condition 
\begin{equation}
(f, g) 
= \int \dd x \, \left[ \calP \calT f(x) \right] \, g(x) 
= \int \dd x \, \calT f(x) \, \calP g(x) 
= \int \dd x \, f^*(x) \, \calP g(x) \,,
\end{equation}
which is equivalent to our $\calP$ scalar product discussed in 
Sec.~\ref{ortho_prop}. Now, for the case of a discrete spectrum, 
one uses Eq.~\eqref{HPTprop} and obtains~\cite{BeBrReRe2004}
\begin{equation}
(\phi_n, \phi_m) 
= \int \dd x \, \phi_n^*(-x) \, \phi_m(x) 
= \int \dd x \, \phi_n(x) \, \phi_m(x) 
= (-1)^n \, \delta_{nm} \,.
\end{equation}
A scalar product is positive definite if its matrix representation is
positive definite in a complete basis set of functions. 
Let us choose the basis set to be the set of the 
eigenstates $\phi_n(x)$. Then, if we can find an operator 
$\calC$ which has the following properties,
\begin{equation}
\calC \, \phi_n(x) = (-1)^n \, \phi_n(x) \,,
\quad
\left< \left. f \right| g \right>
\equiv \int \dd x \, \left[ \calC \calP \calT f(x) \right] \, g(x)  \,,
\quad
\left< \left. \phi_n \right| \phi_m \right> = \delta_{nm} \,,
\end{equation}
the scalar product $\left< \left. \cdot \right| \cdot \right>$
is positive definite. This defines the $\calC$ operator
whose calculation is discussed in Refs.~\cite{BeBrReRe2004,BeBrChWa2005CP}.

The above formalism is somewhat tied to systems with 
a discrete spectrum of bound states. For the continuum 
eigenstates of the tachyonic Dirac Hamiltonian, the norm 
in the $\calP$ scalar product vanishes 
because for plane-wave states of wave vector $\vec k$,
it is proportional to the expression
\begin{align}
\int \dd^3 x \, \left[ \exp(\ii \, \vec k \cdot \vec x) \right]^* \,
\calP \, \exp(\ii \, \vec k \cdot \vec x) =& \;
\int \dd^3 x \, \exp(-\ii \, \vec k \cdot \vec x) \,
\, \exp(-\ii \, \vec k \cdot \vec x) 
\nonumber\\[2ex]
=& \; (2\pi)^3 \, \delta^3(2 \vec k) = 0 \,,
\qquad
\qquad
\vec k \neq \vec 0 \,.
\end{align}
The physically interesting states have a nonvanishing 
wave vector and thus, zero $\calP$ norm. A possible choice for the 
$\calC$ operator, which is physically consistent 
within the manifold of positive-energy and negative-energy states,
therefore is $\calC = \calP$, as discussed in Sec.~\ref{ortho_prop}.
By ``physically consistent'' we mean that it leads to a 
conserved scalar product within the time evolution for the 
positive- and negative-energy states.

%
%
\section*{Acknowledgments}

This work was supported by the NSF and by the
National Institute of Standards and Technology
(precision measurement grant).

\end{document}